\documentclass{elsart}

\usepackage{natbib}

\usepackage{epsfig}

\usepackage{amssymb}

\def\mbh{{$\mathcal M_{\rm BH}$}}
\def\mbulge{{$\mathcal M_{\rm bulge}$}}

\def\mr{{$M_R$}}
\def\lr{{$L_R$}}

\def\mstar{{$\mathcal M_*$}}

\begin{document}

\begin{frontmatter}



\title{Lensed Quasar Hosts\thanksref{label1}}

\thanks[label1]{Observations presented in this paper were obtained using the
Hubble Space Telescope, operated by the Space Telescope Science Institute under
contract to NASA.}


\author[STScI]{Chien Y. Peng},
\author[Steward]{Chris D. Impey},
 \author[MPIA]{Hans-Walter Rix},
 \author[CfA]{Emilio E. Falco}
 \author[Rutgers] {Charles R. Keeton},
 \author[OSU]{Chris S. Kochanek},
 \author[CfA]{Joseph Leh\'ar}, \&
 \author[CfA]{Brian A. McLeod}

 \address[STScI]{Space Telescope Science Institute, 3700 San Martin Drive,
        Baltimore, MD 21218}
 \address[Steward]{Steward Observatory, Univ. of Arizona, 933 N. Cherry
        Ave., Tucson, AZ 85721} 
 \address[MPIA]{Max-Planck-Institut f\"{u}r Astronomie, K\"onigstuhl 17,
        Heidelberg, D-69117, Germany}
 \address[CfA]{Harvard-Smithsonian Center for Astroph., 60 Garden
        St., Cambridge, MA 02138}
 \address[Rutgers]{Department of Physics \& Astronomy, Rutgers University, 136
	Frelinghuysen Road, Piscataway, NJ 08854}
 \address[OSU]{Department of Astronomy, The Ohio State University, 4055
        McPherson Lab, 140 West 18th Avenue, Columbus, OH 43210}

\begin{abstract}

Gravitational lensing assists in the detection of quasar hosts by amplifying
and distorting the host light away from the unresolved quasar core images.  We
present the results of {\it HST} observations of 30 quasar hosts at redshifts
$1 < z < 4.5$.  The hosts are small in size ($r_e \lesssim 6$ kpc), and span a
range of morphologies consistent with early-types (though smaller in mass) to
disky/late-type.  The ratio of the black hole mass ($M_{BH}$, from the virial
technique) to the bulge mass ($M_{bulge}$, from the stellar luminosity) at $1
\lesssim z \lesssim 1.7$ is broadly consistent with the local value; while
$M_{BH}/M_{bulge}$ at $z \gtrsim 1.7$ is a factor of 3--6 higher than the
local value.  But, depending on the stellar content the ratio may decline at
$z\gtrsim4$ (if E/S0-like), flatten off to 6--10 times the local value (if
Sbc-like), or continue to rise (if Im-like).  We infer that galaxy bulge
masses must have grown by a factor of 3--6 over the redshift range $3\gtrsim
z \gtrsim 1$, and then changed little since $z \sim 1$.  This
suggests that the peak epoch of galaxy formation for massive galaxies is above
$z \sim 1$.  We also estimate the duty cycle of luminous AGNs at $z
\gtrsim 1$ to be $\sim 1\%$, or $10^7$ yrs, with sizable scatter.

\end{abstract}


\end{frontmatter}


\section{Introduction}

Elsewhere in these proceedings the reader can find summaries of previous work
on quasar hosts. We concentrate here on the benefits and challenges of using
gravitational lensing as a technique for measuring the host galaxy, especially
in regimes where traditional direct imaging methods are difficult -- at high
redshift, or when the host is either sub-luminous or compact.  The context for
this work is the coevolution of galaxies and supermassive black holes.  In the
local universe, a tight relation is observed between bulge mass (\mbulge) and
black hole mass (\mbh) measured with stellar kinematics \citep{Geb00, Fer00}.
We now extend the relation to $z \gtrsim 1$ using the virial technique to
estimate $M_{BH}$ (e.g. Kaspi et al., 2000; Vestergaard \& Peterson, 2006),
and the stellar luminosity to estimate \mbulge\ \citep{Kor95,Mag98}.  The
existence, slope, and scatter of a \mbh/\mbulge\ relation at high redshift can
be used to analyze the relative growth rates of galaxies and their (presumably
ubiquitous) central engines.

The CfA-Arizona Space Telescope Lens Survey (CASTLES) is a project to image
all known, multiply-imaged, quasars in a homogeneous set of optical and near
infrared passbands using the {\it Hubble Space Telescope} ({\it HST}). With
the number of lens systems now near 100, data are in hand for 80 targets.  The
observations are shallow, 1-2 orbits per filter, but the excellent surface
brightness sensitivity of the {\it HST} leads to a host detection in most
cases.  The overall CASTLES project is described by \citet{Fal01}; early
results in the lensed host search are given by \citet{Rix01} in the same
conference proceedings; and detections and models of the hosts of several
individual quasars have been published \citep{Imp98,Koc00,Kee00}.

Gravitational lensing is a large and growing field of astrophysics so only the
rudiments can be given here; a number of excellent book and reviews are
available for the full formalism and diverse applications
\citep{Bla92,Sch92,Cla02}. About one in 500 quasars has a sight-line passing
close enough to the central potential of a massive galaxy for multiple image
formation. It has taken surveys of tens of thousands of radio and optically
selected quasars to yield the sample of $\sim$100 objects (see the CASTLES web
site\footnote{http://www.cfa.harvard.edu/castles} and that of Li\'ege
group\footnote{http://vela.astro.ulg.ac.be/themes/extragal/gravlens}). Even
though adaptive optics techniques from the ground are improving, stable point
spread functions, obtained with {\it HST}, are still essential for reliable
modeling.


In gravitational lensing, the AGN is magnified into multiple images, but
remains unresolved, whereas the extended light from the host galaxy maps into
arcs or Einstein rings (ER).  A lens model is needed to extract the full
information content of the lensed host light.  In principle, a typical ER
having a radius $\sim$ 1 arcsec means that {\it HST} imaging potentially
yields 50-100 resolution elements in the host galaxy in the deep images from
the survey.

\begin {figure}
\hskip 1.2cm
\vskip -0.1in
\vbox{
\hbox{
\hskip +1.0in
\psfig{file=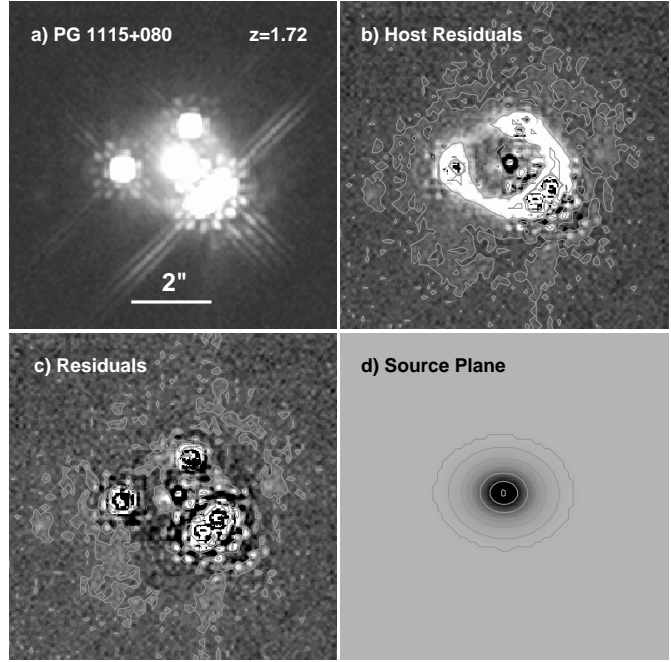,height=3.5truein,angle=0}
}
}
\vskip -0.1in

\caption {An example of the lens modeling technique.  (a) Original NIC2 image
of PG~1115+080 ($z_{\mbox QSO} = 1.72$).  (b) The host galaxy Einstein ring,
after removing the best fit lensing galaxy and quasar point sources.  (c) The
best fitting residuals.  (d) The parametric model of the host galaxy in the
source plane.}
\label{fig:pg1115}

\end {figure}


\section{Modeling}

The image modeling (Fig.~\ref{fig:pg1115}) uses a custom-built program called
LENSFIT (Peng et al.  2006, in prep.), which is based on a methodology that
has been well-tested with the GALFIT algorithm \citep{Pen02}.  The model for
the light profiles of the host and foreground (lens) galaxy uses a S\'ersic
model with a concentration index $n$ that is often used to quantify the gross
morphology of galaxies (e.g.  $n=1$ for late-type, while $n=4$ for
early-type).  Both the quasar point source and the host galaxy light profile
are propagated through the lens model to produce the image distortion, and
multiple images.  External shear is included to model the tidal influence due
to neighbors.  All the parameters are simultaneously varied to reduce the
$\chi^2$ on a pixel-by-pixel basis.  The models are often very robust in well
resolved systems ($\theta \ge 1$ arcsec), due to the spatial separation
between the host and the lens, and their different shapes.


\begin {figure}
\hskip 1.2cm
\vskip -0.1in
\vbox{
\hbox{
\hskip +1.0in
\psfig{file=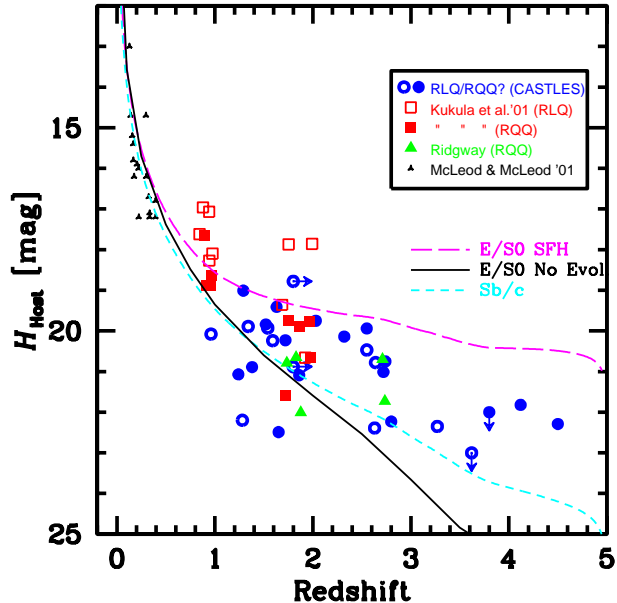,height=3.3truein,angle=0}
}
}
\vskip -0.1in

\caption {The (deprojected) host galaxy luminosity with redshift.  Open
	points: radio loud quasar hosts.  Solid points: radio quiet or unknown
	(most likely radio quiet).  The stellar evolution tracks correspond to
	$z_{form} = 5$, all normalized to $L^*$ by $z=0$.  The E/S0 and Sb/c
	tracks have an initial burst of stars, followed by star formation
	rates that produce colors of E/S0 and Sb/c type galaxies by $z=0$.
	The ``No Evolution'' model corresponds to a redshifted spectrum of a
	$z=0$ E/S0 galaxy.
}
\label{fig:radio}

\end {figure}


\section{Results}

Here we present a summary of our findings, which are detailed along with a
description of our analysis techniques elsewhere (Peng et al. 2006, in prep.).
We select on lensing geometry size ($\theta \gtrsim 0.7$ arcsec), which does
not a priori bias the intrinsic AGN luminosity selection, or the host
luminosities.  Therefore, we expect the sample of the AGNs to be randomly
drawn from the AGN luminosity function, where the lower limits are determined
by various lensing search programs.  The heterogeneity of different surveys,
however, will not affect our primary conclusions about the relationships
between \mbh\ and \mbulge, since both quantities are measured in the {\it
same} objects.

\subsection {General Properties}

Figure \ref{fig:radio} shows the $H$-band host galaxy luminosities (lensing
distortion removed) versus redshift.  Overall, the host luminosities from
CASTLES appear to agree well with non-lensing studies \citep{Kuk01,Rid01}.
The host luminosities range from 1 to 20 times $L^*_V$ today, while the AGNs
are 0 to 3 magnitudes brighter than the host in restframe $B$ to $V$ band.
Despite their brightnesses, the host galaxies appear to be fairly small in
size (typical $r_e \lesssim 6$ kpc) for their central AGNs.  As we shall see
later, when coupled with information about their \mbh, both the host
luminosities and sizes lead to the conclusion that the bulges may be
undermassive compared to present-day normal galaxies.  Lastly, the S\'ersic
index values suggest that while a number of quasar hosts at $z\gtrsim 1.5$
have steep central concentrations consistent with the presence of a bulge,
many (30\% -- 50\%) also have low S\'ersic values ($n \le 2$) more analogous
to later-type galaxies today.  Even those galaxies with high S\'ersic indices
may not qualify as bona fide -- fully formed and passively evolving --
ellipticals, given their small sizes and black hole masses.

{\it Radio-loud (RLQ) vs. Radio-quiet (RQQ) hosts}\ \ \ \ \ Quantifying
differences between RLQ and RQQ hosts has historically been controversial.
Figure~\ref{fig:radio} shows that there is not a clear difference in the host
luminosities between RLQ (open) and RQQ (closed) AGNs in the lensing sample.
The diverse selection criteria from disjointed surveys are, however, hard to
quantify.  It is worth to keep in mind, however, that in a study by
\citet{Kuk01} RLQs were drawn from the rare and extreme radio-loud sources
which may require atypically large BHs to produce.  Consequently, the finding
of luminous hosts in those RLQs may reflect a correlation between
\mbh\ and \mbulge\ at high redshifts.  The issue of radio correlation with
host properties remains unsettled.

{\it AGN Duty Cycle}\ \ \ \ \ We can estimate the duty cycle of nuclear
activity for each object with a host detection.  A rough estimate of the duty
cycle is: $D \sim\Phi_Q(L_{QSO},z)/\Phi_G(L_{Gal},z)$, where $\Phi_Q$ and
$\Phi_Q$ are the luminosity functions of quasars and galaxies, respectively,
appropriate to a given redshift.  At $z\gtrsim 1$, the median duty cycle is
1\%, or $10^7$ years, with a sizable scatter.

\subsection {Black Hole vs. Bulge Evolution} 

Based on quasar and host luminosities we can study the \mbh\ vs. bulge
properties at $z\gtrsim 1$ (see Peng et al. 2005 for details), where \mbh's
are obtained using the virial technique \citep {Kas00,Ves06}.

\begin {figure}
\hskip 1.2cm
\vskip -0.1in
\vbox{
\hbox{
\hskip +0.4in
\psfig{file=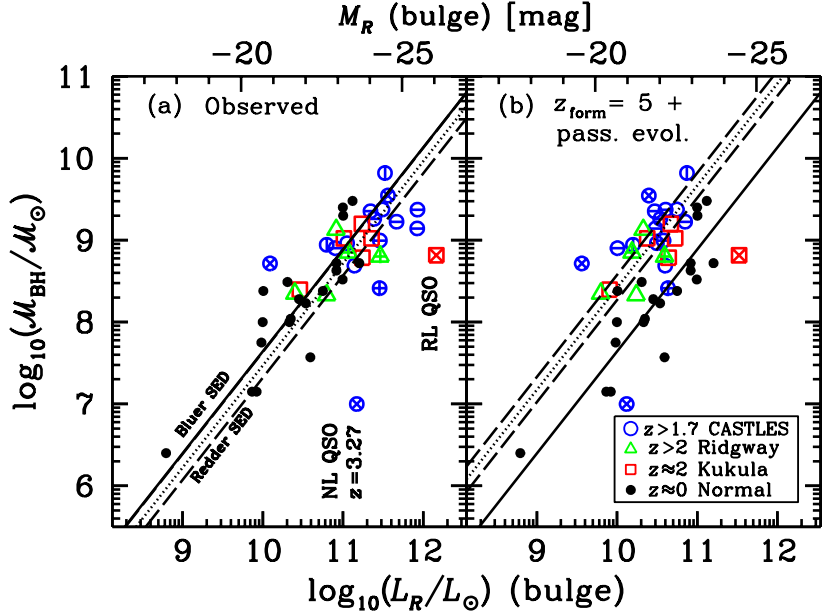,height=4.5truein,angle=-90}
}
}
\vskip -0.1in

\caption {The relationship of the black hole mass, \mbh, vs.  bulge absolute
      luminosity (\lr, bottom axis; \mr, top axis), at low $z$ (solid round
      points) and $z\gtrsim 1.7$ (open points).  Solid lines:  fit to
      $z\approx0$ solid points.  All open points assume a modern-day Sbc-type
      SED for $K$-correction and their average is represented by dotted lines.
      Dashed lines illustrate assumptions of bluer(Im, left)/redder(E, right)
      SEDs.  Open circle:  gravitationally lensed quasar hosts.  Open
      triangles:  \citet{Rid01}.  Open squares:  \citet{Kuk01}.  Vertical line
      in points:  a possible lower limit in \mbh\ due to AGNs being broad
      absorption line QSOs.  Criss-crossed points:  potential problem with
      lens identification, host detection, radio-loud quasar, or narrow line
      AGN.  Panel ({\it a}):  The observational data.  Panel ({\it b}):  The
      same data in {\it a}, but the open points are shifted horizontally by
      assuming that the hosts evolve {\it passively} with $z_f=5$ by
      $d$\mr/$dz$ = $-0.8$ mag.  See also \citet{Pen05} for details.
}

\label{fig:highz}
\end {figure}

{\it $1.7 \lesssim z\lesssim 4.5$}\ \ \ \ \ Fig.~\ref{fig:highz} shows the
\mbh\ vs. the restframe $R$-band bulge luminosity (\lr) for the host galaxies
at $1.7 \lesssim z \lesssim 4.5$.  Determining \lr\ requires $K$-corrections,
computed using an Sbc SED; the dashed lines shows the small systematic effect
of using an E/S0 (right) or Im (left) SED.  It is clear that a correlation
between \mbh\ and bulge luminosity was already present at a lookback time of
10--12 Gyr.  Remarkably, the high-redshift hosts appear to lie on the {\em
same} relation as $z=0$ normal galaxies, implying that the high-$z$ hosts are
{\em undermassive} in comparison.  To explain why, Fig.\ 3b shows the host
luminosities after we account for passive evolution of $dM_R/dz = -0.8$ mag;
specifically, we hold $M_{BH}$ fixed and shift the color points in Fig.\ 3a to
the left.  Now the $z\gtrsim 1.7$ hosts are displaced from the local host
relation (solid line) by a factor of 3--6 in luminosity, which translates into
a mass deficit of a factor of 3--6 in the quasar hosts compared to local
galaxies with the same \mbh.

\begin {figure}
\hskip 1.2cm
\vskip -0.1in
\vbox{
\hbox{
\hskip +0.4in
\psfig{file=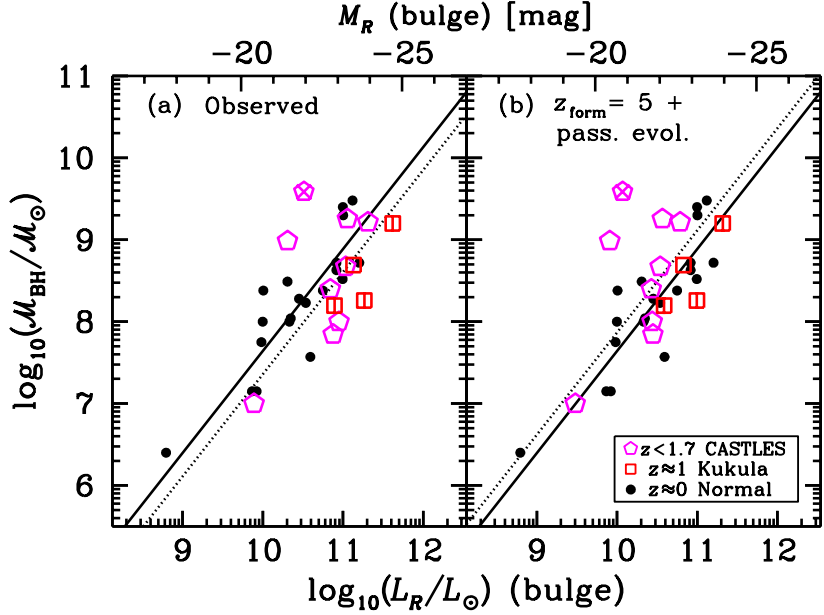,height=4.5truein,angle=-90}
}
}
\vskip -0.1in

\caption {The same diagram as Figure \ref{fig:highz}, except for redshift of
	$1 \lesssim z \lesssim 1.7$ quasar hosts.  The vertical line in the
	square points indicates that the AGN either has a strong narrow
	Mg~{\sc ii} line component or is strongly absorbed in the wings,
	causing a potentially low \mbh\ estimate.  The dotted line is
	displaced from the solid line, representing the local \mbh-\lr\
	relation, by $-0.5$ (Fig. {\it a}) and $+0.5$ (Fig. {\it b})
	magnitude. Note the very slight bias between the non-lensed and lensed
	datasets, which might be explained by the difference in the median
	redshift of $\left<z\right>_{\mbox{med}} = 1.45$ for the lens sample
	and $\left<z\right>_{\mbox{med}} = 0.94$ for the non-lenses.  However,
	three of the non-lensed data points may also have a lower limit on the
	\mbh\ estimate, as noted above.}
\label {fig:lowz}

\end {figure}

{\it $1\lesssim z \lesssim 1.7$}\ \ \ \ \ In contrast, by $z\approx 1$,
Fig.~\ref{fig:lowz} shows that the mass deficit of the hosts is mostly reduced
(to within a factor of unity in mass), after accounting solely for passive
evolution.  Thus, massive bulges that correspond to luminous E/S0 galaxies
today may have been nearly assembled by $z\approx 1$.

\begin {figure}
\hskip 1.2cm
\vskip -0.1in
\vbox{
\hbox{
\hskip +0.12in
\psfig{file=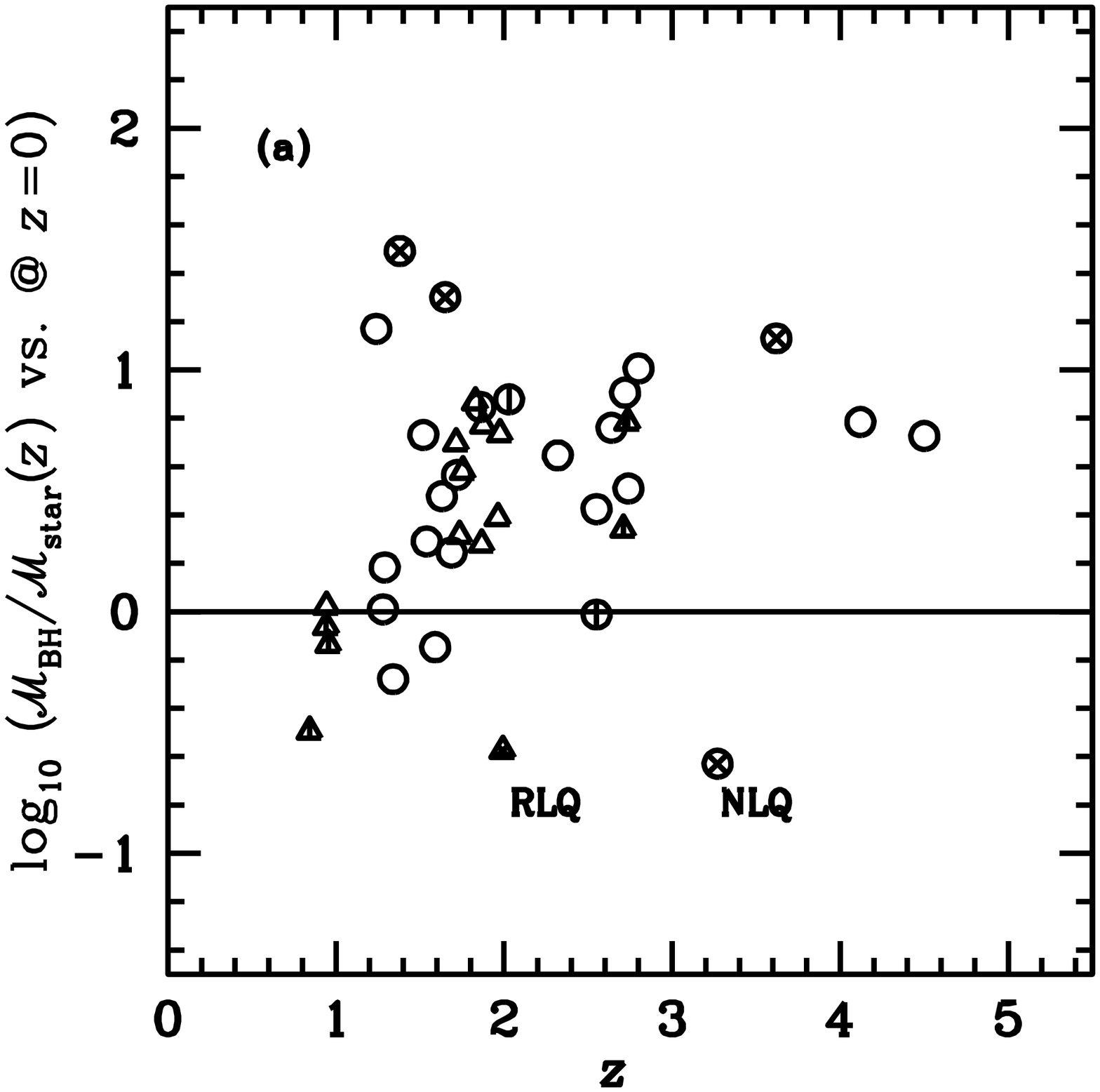,height=2.4truein,angle=0}
\hskip +0.16in
\psfig{file=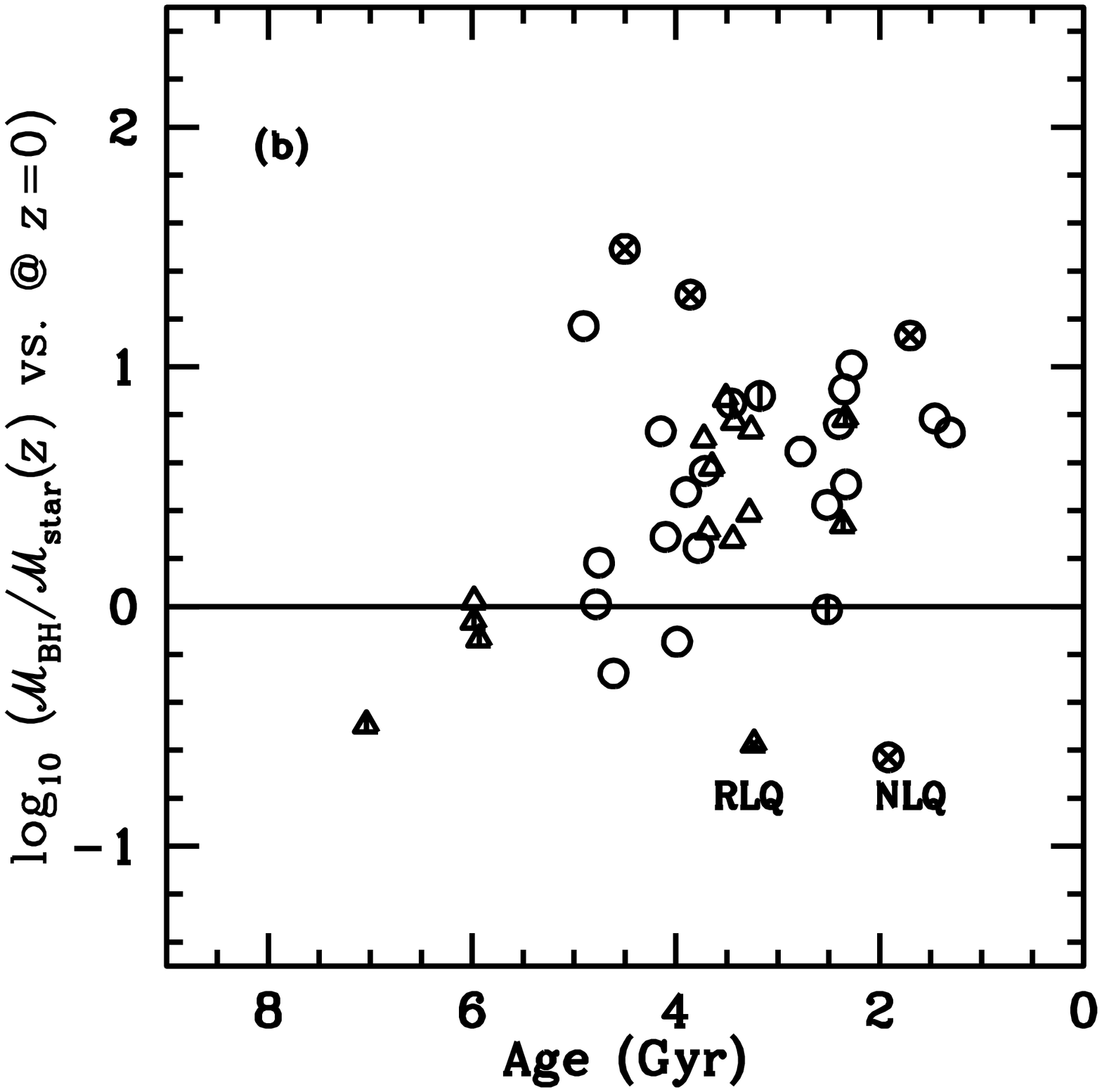,height=2.4truein,angle=0}
}
}
\vskip -0.1in

\caption {The growth of the \mbh/\mstar\ ratio as a function of ({\it a})
	redshift and ({\it b}) age of the universe in Gyrs.  Circles are
	gravitational lens data points, while triangles are from direct
	imaging of hosts using {\it HST} NICMOS $H$-band \citep{Rid01,Kuk01}.
	Point styles are the same as Fig.~\ref{fig:highz}.  The \mbh/\mstar\
	ratio appears to rise quickly beyond $z\approx 1$ and may slow, and
	perhaps flatten, to a factor of $6-10\times$ local value by $z\approx
	3$.  A fading rate of $dM_R/dz=-0.8$ is assumed here (passive
	evolution since $z_{form} = 5$).
}
\label{fig:growth}
\end {figure}

By requiring that the hosts evolve onto the local \mbh\ vs. \lr\ of
Figs.~\ref{fig:highz} and \ref{fig:lowz}, we can illustrate a growth in the
\mbh/\mbulge\ ratio with redshift relative to today, shown in
Figure~\ref{fig:growth}.  We find that the ratio of \mbh/\mbulge\ increases
roughly to a factor of 10 higher than today (assuming an Sbc-type SED) out at
$z=4$.  This conclusion depends somewhat on the assumption of the SED and
evolutionary history: the ratio at earlier times would be higher than shown
for a bluer SED than Sb/c, or for a faster fading rate than passive evolution.


\section{Conclusions}

Detailed modeling of 30 well-observed systems from a total sample of 80 lensed
quasars has provided new insights into the properties of host galaxies at $1 <
z < 4.5$.  About half have S\'ersic model fits indicative of early type
galaxies.  However, combined with their small sizes of $r_e < 6$ kpc,
luminosities, and \mbh, it appears that luminous, fully-formed, ellipticals
are in a minority as hosts of luminous quasars at $z\gtrsim 2$. No difference
is seen between the luminosities of radio-loud and radio-quiet quasars in the
sample, with a caveat on sample selection.  Even at $z \gtrsim 2$, the host
galaxies follow nearly the same relationship between \mbh\ and luminosity as
at low redshifts, but the bulges must gain in mass by a factor of 3-6 between
$1\lesssim z \lesssim 4.5$. However, by $z\approx1$, the mass deficit is
mostly gone.  Thus massive bulges at $z\approx 1$ may be consistent with being
passively evolving, or may still grow by at most a factor of 1.  Our estimate
of the AGN duty cycle is $\approx 1\%$, or $10^7$ years.  Ongoing work
includes using color information to constrain the host star formation
histories, obtaining sub-millimeter data to measure star formation rate, and
characterizing detailed host morphology with deeper {\it HST} imaging.
 



\end{document}